\documentclass[sigconf]{acmart}

\usepackage{array}
\newcommand{\PreserveBackslash}[1]{\let\temp=\\#1\let\\=\temp}
\newcolumntype{C}[1]{>{\PreserveBackslash\centering}p{#1}}
\newcolumntype{R}[1]{>{\PreserveBackslash\raggedleft}p{#1}}
\newcolumntype{L}[1]{>{\PreserveBackslash\raggedright}p{#1}}
\usepackage{booktabs}  

\usepackage{amsfonts}
\usepackage{amsmath}
\usepackage{balance}  
\usepackage{graphicx}
\usepackage{enumitem}
\usepackage{multirow}
\usepackage[ruled]{algorithm}
\PassOptionsToPackage{noend}{algpseudocode}
\usepackage{algpseudocode}
\usepackage{subfigure}
\usepackage{url}

\usepackage{xcolor,colortbl}
\colorlet{shadecolor}{gray!15}
\usepackage{makecell}
\newcommand\newtoprule{\Xhline{.10em}}
\newcommand\newmidrule{\Xhline{.05em}}
\newcommand\newbottomrule{\Xhline{.10em}}

\renewcommand{\algorithmiccomment}[1]{\bgroup\hfill//~#1\egroup}

\algnewcommand\algorithmicswitch{\textbf{switch}}
\algnewcommand\algorithmiccase{\textbf{case}}
\algnewcommand\algorithmicassert{\texttt{assert}}
\algnewcommand\Assert[1]{\State \algorithmicassert(#1)}%
\algdef{SE}[SWITCH]{Switch}{EndSwitch}[1]{\algorithmicswitch\ #1\ \algorithmicdo}{\algorithmicend\ \algorithmicswitch}%
\algdef{SE}[CASE]{Case}{EndCase}[1]{\algorithmiccase\ #1}{\algorithmicend\ \algorithmiccase}%
\algtext*{EndSwitch}%
\algtext*{EndCase}%
\algnewcommand\And{\textbf{and} }

\newcommand{\gnn}{ED-GNN}

\AtBeginDocument{%
  \providecommand\BibTeX{{%
    \normalfont B\kern-0.5em{\scshape i\kern-0.25em b}\kern-0.8em\TeX}}}

\copyrightyear{2021}
\acmYear{2021}
\setcopyright{acmcopyright}
\acmConference[SIGMOD'21] {Proceedings of the 2021 International Conference on Management of Data}{June 18--27, 2021}{Virtual Event, China}
\acmBooktitle{Proceedings of the 2021 International Conference on Management of Data (SIGMOD'21), June 18--27, 2021, Virtual Event, China}
\acmPrice{15.00}
\acmDOI{10.1145/3448016.3457328}
\acmISBN{978-1-4503-8343-1/21/06}

\begin{document}
\fancyhead{}

\title{Medical Entity Disambiguation Using Graph Neural Networks}

\author{Alina Vretinaris$^{1*}$, Chuan Lei$^2$, Vasilis Efthymiou$^{3*}$, Xiao Qin$^2$, Fatma \"Ozcan$^{4*}$}
\affiliation{
    $^1$IBM Germany, Ehningen, Baden-Württemberg, Germany\\
    $^2$IBM Research - Almaden, 650 Harry Road, San Jose, CA 95120\\
    $^3$FORTH - Institute of Computer Science, Heraklion, Crete, Greece\\
    $^4$Google, 1600 Amphitheatre Parkway, Mountain View, CA, 94043
}
\email{alina.vretinaris|chuan.lei|xiao.qin@ibm.com, vefthym@ics.forth.gr, fozcan@google.com}

\thanks{*Work done while at IBM Research.}

\renewcommand{\authors}{Alina Vretinaris, Chuan Lei, Vasilis Efthymiou, Xiao Qin, and Fatma \"Ozcan}

\begin{abstract}
Medical knowledge bases (KBs), distilled from biomedical literature and regulatory actions, are expected to provide high-quality information to facilitate clinical decision making. Entity disambiguation (also referred to as entity linking) is considered as an essential task in unlocking the wealth of such medical KBs. However, existing medical entity disambiguation methods are not adequate due to word discrepancies between the entities in the KB and the text snippets in the source documents. Recently, graph neural networks (GNNs) have proven to be very effective and provide state-of-the-art results for many real-world applications with graph-structured data. In this paper, we introduce \gnn\ based on three representative GNNs (GraphSAGE, R-GCN, and MAGNN) for medical entity disambiguation. We develop two optimization techniques to fine-tune and improve \gnn. First, we introduce a novel strategy to represent entities that are mentioned in text snippets as a query graph. Second, we design an effective negative sampling strategy that identifies hard negative samples to improve the model's disambiguation capability. Compared to the best performing state-of-the-art solutions, our \gnn\ offers an average improvement of 7.3\% in terms of F1 score on five real-world datasets.
\end{abstract}



\maketitle

\section{Introduction}
\label{sec:intro}

Recent years have witnessed the rapid growth in medical knowledge bases (KBs), curated from healthcare data, such as clinical resources, electronic health records, and lab tests. Tremendous effort has been put into developing automated medical KB construction~\cite{10.5555/1873781.1873813,Wright2019NormCoDD} and completion~\cite{10.5555/2832415.2832507,nguyen-etal-2018-novel}. Existing systems often face one major challenge, \textit{entity disambiguation} (ED): how to map entity mentions in text snippets from medical source documents to their corresponding entities in a medical KB.

Text snippets in healthcare data are often collected from heterogeneous data sources. Discrepancies arise for many reasons, including acronyms, abbreviations, typos and colloquial terms. As a result, text snippets may deviate significantly from the canonical descriptions of the entities in the KB that they refer to. For example, an editorial staff member may mention \textit{``renal disorder''} or \textit{``kidney disease''} in a text snippet, with the intention to refer to the entity that is defined as \textit{``nephrosis''} in the KB. Similarly, \textit{``cah''} in a text snippet may refer to the entity defined as \textit{``chronic active hepatitis''}. Such discrepancies make it difficult to link textual entity mentions to the intended entities in a KB, introducing noise, duplicates, and ambiguity, eventually decreasing the value of the KB. 

While early works often relied on rule-based~\cite{DBLP:journals/jamia/TikkS10,DBLP:conf/sigmod/HuaZZ15,DBLP:journals/artmed/KorkontzelosPDA15} and dictionary-based approaches~\cite{DBLP:journals/jamia/SavovaMOZSSC10,DBLP:journals/bmcbi/TseytlinMLCCJ16}, more recent state-of-the-art ED solutions rely on machine learning methods. In particular, deep learning (DL) methods~\cite{DBLP:journals/tkde/ShenHWYY18,Dai:2018:FCL:3183713.3196907,Wright2019NormCoDD,DBLP:conf/sigmod/GovindKCMNLSMBZ19} are commonly used due to their powerful feature abstraction and generalization capabilities. A recent study~\cite{10.1145/3183713.3196926} of various DL-based methods for entity matching, concluded that they significantly outperform other solutions (e.g., \cite{DBLP:conf/sigmod/GovindKCMNLSMBZ19}) for textual entity matching. However, existing DL methods either resolve mentions only relying on textual context information from the surrounding words~\cite{chisholm-hachey-2015-entity,Dai:2018:FCL:3183713.3196907,Wright2019NormCoDD}, or merely use entity embeddings for feature extraction and rely on other modules for ED~\cite{DBLP:journals/tkde/ShenWH15,DBLP:journals/tkde/ShenHWYY18,Dai:2018:FCL:3183713.3196907}. They do not fully exploit the structural information in text snippets and KBs.

Recently, graph representation learning has emerged as an effective approach to learn vector representations for graph-structured data. Graph Neural Networks (GNNs)~\cite{DBLP:conf/iclr/KipfW17,DBLP:conf/nips/HamiltonYL17,HAN} have shown promising results in various representation learning tasks on KBs, including link prediction, node classification, as well as node clustering. The foundation of GNNs is a powerful spatial invariant aggregation function that learns how to aggregate rich structural and semantic information from each node’s neighborhood to generate node embeddings. Motivated by the observation that entity mentions in a text snippet are likely to share similar or relevant context, we represent these entity mentions as a query graph to capture their interdependence. Then, we model ED as a graph matching problem and propose a simple architecture, \gnn, which not only collectively learns the contextual information and structural interdependence of entity mentions in the given text snippets, but also captures discriminative contextual information of entities in a medical KB. We target the medical domain because medical KBs contain deep and fine-grained knowledge, which is reflected by their rich hierarchical structure and vocabularies that can be utilized by our \gnn. Note that \gnn\ could be applied to other domain-specific or cross-domain KBs as well, if they contain similar contextual or structural characteristics as the medical ones.

We propose two optimizations for \gnn\ to further improve its disambiguation capability. First, after constructing a \emph{query graph} (representing the entity mentions in a text snippet), \gnn\ augments this graph with domain knowledge from the medical KB. Consider the text snippet \textit{``Aspirin can cause nausea indicating a potential ARF, nephrotoxicity, and proteinuria''}. The abbreviation \textit{``ARF''} is a mention that could refer to the entities \textit{``acute renal failure''} or \textit{``acute respiratory failure''} in the medical KB. Leveraging the domain knowledge from the medical KB (i.e., that \textit{``nephrotoxicity''} and \textit{``proteinuria''} are adverse effects of Aspirin), \gnn\ understands that \textit{``ARF''} is in the context of Aspirin's adverse effects. Hence, \textit{``acute renal failure''} is identified as the matching entity, even though the abbreviation of \textit{``acute respiratory failure''} is also \textit{``ARF''}.

Second, \gnn\ is equipped with an effective negative sampling strategy, which challenges \gnn\ to learn from {\it difficult} samples  to improve the model's disambiguation capability. Assume that we have picked up (\textit{``ARF''}, \textit{``acute renal failure''}) as a positive training example. Following convention~\cite{DBLP:conf/icde/ZhangYSC19}, we sample negative examples by replacing \textit{``acute renal failure''} from the above positive example. Then (\textit{``ARF''}, \textit{``chronic renal failure''}) is a difficult negative sample as the lexical similarity between \textit{``chronic renal failure''} and \textit{``acute renal failure''} is high. Another difficult negative sample can be (\textit{``ARF''}, \textit{``gastroenteritis''}) since \textit{``gastroenteritis''} shares several common neighbors with \textit{``acute renal failure''} in the medical KB. 
\gnn\ can more effectively learn from the above negative samples to reach the desired accuracy, compared to the commonly used random negative sampling~\cite{DBLP:conf/iclr/KipfW17} that replaces \textit{``acute renal failure''} with a random entity (e.g., \textit{``fever''}) in the medical KB.

\textbf{Contributions.} We highlight our main contributions as follows:

\begin{itemize}[leftmargin=*]
\item We present \gnn, a novel medical ED solution, based on graph neural networks (GNNs) such as GraphSAGE~\cite{DBLP:conf/nips/HamiltonYL17}, R-GCN~\cite{DBLP:conf/esws/SchlichtkrullKB18}, and MAGNN~\cite{MAGNN}. We model ED as a graph matching problem to leverage such GNNs with a simple architecture. 

\item We develop two optimization techniques to further improve \gnn's disambiguation capability. First, we construct the query graph and augment it with domain knowledge from the medical KB. This helps \gnn\ focus on the right structural information from the query graph for making the matching decisions. Second, we design an effective negative sampling strategy, which provides \gnn\ with harder examples, resulting in more discriminative power for entity disambiguation.

\item We evaluate the effectiveness of \gnn\ on multiple real-world datasets. Our experimental results show that \gnn\ consistently outperforms the state-of-the-art ED solutions in all datasets by up to 16.4\% in F1 score. Furthermore, we evaluate the two optimization techniques in \gnn\ and show that both of them lead to performance improvements.
\end{itemize}

\textbf{Outline}. The rest of the paper is organized as follows. Section~\ref{sec:preliminary} introduces the basic notation, briefly describes a family of GNNs, and overviews the architecture of \gnn. Section~\ref{sec:opt} describes the two optimization techniques designed for \gnn. We present our experiments in Section~\ref{sec:exp}, review related work in Section~\ref{sec:related}, and conclude in Section~\ref{sec:conclusion}.
\section{Background and Architecture}
\label{sec:preliminary}

\begin{definition}(Heterogeneous Graph)
We define a heterogeneous graph as a graph $\mathcal{G}$ = ($\mathcal{V}$, $\mathcal{E}$) associated with a node type mapping function $\phi$ : $\mathcal{V}\mapsto\mathcal{T}$ and an edge type mapping function $\psi$ : $\mathcal{E}\mapsto\mathcal{R}$, where $\mathcal{T}$ and $\mathcal{R}$ denote the sets of node types and edge types, respectively, with $|\mathcal{T}|$ + $|\mathcal{R}| > 2$.
\label{def:HG}
\end{definition}

\begin{figure}[!htb]
\centering
\includegraphics[width=0.8\columnwidth]{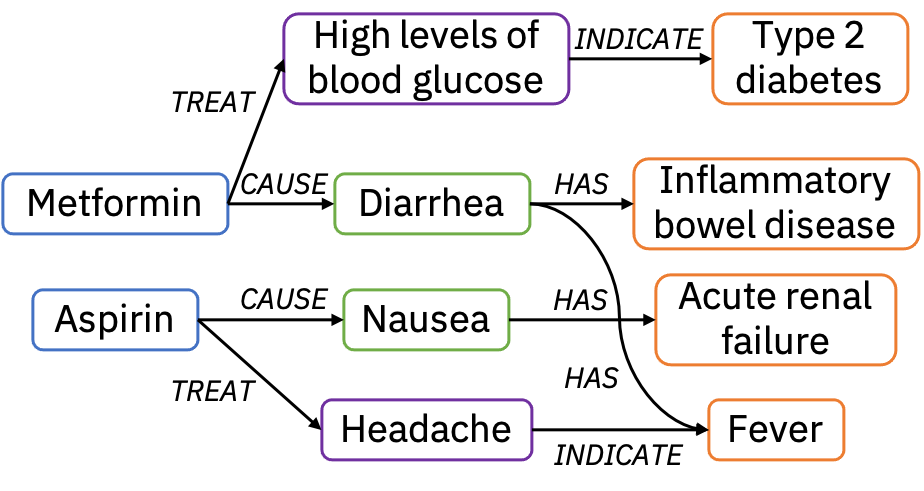} 
\caption{A toy heterogeneous graph (best viewed in color).}
\label{fig:hg}
\end{figure}

Figure~\ref{fig:hg} shows a toy example of a heterogeneous graph constructed from a medical KB. The node types are Drug (blue nodes), AdverseEffect (green nodes), Symptom (purple nodes), and Finding (orange nodes). The edge types are TREAT, CAUSE, INDICATE, as well as HAS. Besides, all these nodes are associated with descriptions (e.g., Aspirin, headache, nausea, and fever). In this work, we model both a medical KB and a text snippet as heterogeneous graphs, such that we cast medical ED as a binary classification problem using the expressive power of heterogeneous GNNs.

\begin{definition}(Heterogeneous Graph Embedding)
Given a heterogeneous graph $\mathcal{G}$ = ($\mathcal{V}$, $\mathcal{E}$), with node attribute matrices $A_{T_i}$ $\in$ $\mathbb{R}^{|\mathcal{V}_{T_i}|\times d_{T_i}}$ for node types $T_i$ $\in$ $\mathcal{T}$, a heterogeneous graph embedding is a $d$-dimensional node representation (a.k.a. embedding) for all $v\in \mathcal{V}$ with $d$ $\ll$ $|\mathcal{V}|$, which captures the network structural and semantic information in $\mathcal{G}$. 
\label{def:hne}
\end{definition}

\begin{table}[!htb]
\small
\centering
\caption{Table of notations.}
\begin{tabular}{|c|p{6cm}|}
\hline
Notation                        & Description \\\hline\hline
$\mathcal{G}$                   & Heterogeneous graph \\\hline
$\mathcal{G}_{\it ref}$         & Knowledge base (reference graph) \\\hline
$\mathcal{V}_{\it ref}$         & The set of nodes in $\mathcal{G}_{\it ref}$ \\\hline
$\mathcal{E}_{\it ref}$         & The set of edges in $\mathcal{G}_{\it ref}$ \\\hline
$v^r$                           & A node in $\mathcal{V}_{\it ref}$ \\\hline
$\mathcal{G}_{\it qry}$         & Query graph \\\hline
$\mathcal{V}_{\it qry}$         & The set of nodes in $\mathcal{G}_{\it qry}$ \\\hline
$\mathcal{E}_{\it qry}$         & The set of edges in $\mathcal{G}_{\it qry}$ \\\hline
$v^{q}$                         & A node in $\mathcal{V}_{\it qry}$ \\\hline
$\mathbf{h}_v^{\text{attr}}$    & Initial node feature \\\hline
$\mathbf{h}_v$                  & Hidden state (embedding) of node $v$ \\\hline
$P$                             & A metapath \\\hline
$\mathcal{P}$                   & The set of metapaths \{$P_1$, $P_2$,$\cdots$, $P_M$\} \\\hline
{$P(u, v)$}                     & A metapath instance connecting nodes $u$ and $v$ \\\hline
$\mathcal{N}_v$                 & The set of neighbors of node $v$ \\\hline
$\mathcal{N}^{P}_v$           & The set of neighbors of node $v$ based on $P$ \\\hline
\end{tabular}
\label{tab:notation}
\end{table}

\subsection{Graph Neural Networks}
\label{sec:preliminary:gnn}

In recent years, Graph Neural Networks (GNNs) have been intensively studied and shown effective for various graph mining and analytical tasks, including node classification, link predication, and graph matching. Their ability to combine structural information and semantic features is essential to our ED task. In \gnn, we employ three representative approaches, including GraphSAGE~\cite{DBLP:conf/nips/HamiltonYL17}, R-GCN~\cite{DBLP:conf/esws/SchlichtkrullKB18} and MAGNN~\cite{MAGNN}. GraphSAGE is a seminal message-passing GNN, which employs the general notion of aggregator functions for efficient generation of node embeddings. R-GCN is a relation-aware graph convolutional network which handles $k$-hop message-passing over heterogeneous KBs. MAGNN is the state-of-the-art metapath-based GNN that supports heterogeneous KBs and learns subtle contextual structures in KBs using semantic-aware neighbor aggregation with composite relations. All three GNNs are implemented using Deep Graph Library~\cite{wang2019dgl} on top of PyTorch~\cite{pytorch}. This makes \gnn\ lightweight and easy to adapt to new KBs. Note that other GNNs can be plugged into our architecture as well. Table~\ref{tab:notation} summarizes the notations used in these three GNNs. 

{\bf GraphSAGE.} GraphSAGE~\cite{DBLP:conf/nips/HamiltonYL17} leverages node features (e.g., text descriptions/labels associated with nodes) in order to learn an embedding function that generalizes to unseen nodes. By incorporating node features, GraphSAGE simultaneously learns the topological structure of each node’s neighborhood as well as the distribution of node features in the neighborhood. Formally, the $k$-th layer of GraphSAGE is:
\begin{flalign}
\begin{aligned}
\mathbf{h}^{k}_{\mathcal{N}_v} & = \mathrm{AGGREGATE}({\mathbf{h}^{k-1}_{u},\forall u\in \mathcal{N}_v}),\\
\mathbf{h}^{k}_{v} & = \sigma(\mathbf{W}^k \cdot [\mathbf{h}^{k-1}_v||\mathbf{h}^k_{\mathcal{N}_v}]),
\label{eq.graphsage}
\end{aligned}
\end{flalign}
where $\sigma$ is an activation function and $\mathbf{W}^k$ is a set of weight matrices, $\forall k\in$ $\{$1, ..., $K\}$, which are used to propagate information between different layers of the model. The intuition behind Equation~\ref{eq.graphsage} is that at each layer, nodes aggregate information
from their local neighbors, and as this process iterates, nodes incrementally gain more and more
information from further reaches of the graph.

{\bf R-GCN.} Unlike GraphSAGE that only considers the node-wise connectivity in a graph and ignores edge labels such as the relations in KBs, R-GCN distinguishes different neighbors with relation-specific weight matrices. In the $k$-th convolutional layer, each representation vector is updated by accumulating the vectors of neighboring nodes through a normalized sum: 
\begin{equation}
\mathbf{h}^{(k)}_{v} = \sigma(\mathbf{W}_0^k\mathbf{h}_v^{k-1} + \sum_{r\in\mathcal{R}}\sum_{u\in \mathcal{N}_v^r} \frac{1}{c_{v,r}}\mathbf{W}_r^k\mathbf{h}_u^{k-1}),
\label{eq.rgcn}
\end{equation}
where $\mathbf{W}_0^k$ is the weight matrix for the node itself,  $\mathbf{W}_r^k$ is used specifically for the neighbors having relation $r$, i.e., $\mathcal{N}_v^r$, $\mathcal{R}$ is the relation set, and $c_{v,r}$ is used for normalization. Intuitively, different edge types use different weights and only edges of the same relation type $r$ are associated with the same projection weight $\mathbf{W}_r^k$.

{\bf MAGNN.} MAGNN aggregates a node $v$’s representation from $\mathcal{N}_v^\mathcal{P}$ (i.e., the metapath-aware neighborhood) and the nodes in between, by encoding the metapath instances through a relational rotation encoder. To elaborate, we introduce the following definitions from~\cite{MAGNN}.

\begin{definition}(Metapath)
A metapath $P$ in a heterogeneous graph $\mathcal{G}$
is a path in the form of $A_1$ $\overset{R_1}{\rightarrow}$ $A_2$ $\overset{R_2}{\rightarrow}$ $\cdots$ $\overset{R_m}{\rightarrow}$ $A_{m+1}$ (abbreviated as $A_1A_2$$\cdots$$A_{m+1}$), where $A$ and $R$ are node types and edge types in $\mathcal{G}$, respectively.
\label{def:metapath}
\end{definition}


\begin{definition}(Metapath-based Neighbors)
Given a metapath $P$ of a heterogeneous graph, the metapath-based neighbors $\mathcal{N}^P_v$ of a node $v$ are defined as the set of nodes that connect with node $v$ via metapath instances of $P$.
\end{definition}

For example, Drug-AdverseEffect-Finding (DAF) is a metapath representing that drugs cause adverse effects, and these adverse effects can be described by findings. Given the metapath DAF, \textit{``Fever''} and \textit{``Diarrhea''} constitute the metapath-based neighbors of \textit{``Metformin''} in Figure~\ref{fig:hg}. These nodes are connected with \textit{``Metformin''} via the metapath instance \textit{``Metformin-Diarrhea-Fever''}\footnote{Note that metapath-based neighbors are not limited to 1-hop neighbors.}.

As defined in~\cite{MAGNN}, during the intra-metapath aggregation, each target node extracts and combines information from the metapath instances connecting the node with its metapath-based neighbors. The intra-metapath aggregation layer is formally defined as:
\begin{flalign}
\begin{aligned}
e^P_{vu} & = \mathrm{LeakyReLU}(a^\intercal_P \cdot [\mathbf{h}_v||\mathbf{h}_{P(u, v)}]),\\
\alpha^P_{vu} & = \frac{\mathrm{exp}(e^P_{vu})}{\sum_{s\in \mathcal{N}^P_v} \mathrm{exp}(e^P_{vs})},\\
\mathbf{h}_v^P & = \sigma(\sum_{u\in \mathcal{N}^P_v} \alpha^P_{vu}\cdot \mathbf{h}_{P(v,u)}),
\end{aligned}
\end{flalign}
where $\mathbf{h}_{P(u, v)}$ represents all the node features along a metapath instance, $a_P$ is the parameterized attention vector for the metapath $P$, and $\alpha^P_{vu}$ is the normalized importance weight for all $u\in\mathcal{N}_v^P$. Finally, the intra-metapath output goes through an activation function $\sigma(\cdot)$. In this way, MAGNN captures the structural and semantic information of heterogeneous graphs from both neighbor nodes and the metapaths between the target node and its neighbors.

After aggregating the node and edge information within each metapath, MAGNN uses an inter-metapath aggregation layer with the attention mechanism to fuse latent vectors of the node $v$ obtained from multiple metapaths into final node embeddings. The inter-metapath aggregation layer is formally defined as:
\begin{flalign}
\begin{aligned}
e_{P_i} & = q^\intercal_A\cdot s_{P_i},\\
\beta_{P_i} & = \frac{\mathrm{exp}(e_{P_i})}{\sum_{P\in\mathcal{P}_A}\mathrm{exp}(e_p)},\\
\mathbf{h}_v^{\mathcal{P}_A} & = \sum_{P\in\mathcal{P}_A} \beta_P\cdot \mathbf{h}_v^P,\\
\end{aligned}
\end{flalign}
where $s_{P_i}$ denotes the summarized metapath $P_i\in\mathcal{P}$ by averaging the transformed metapath-specific node vectors for all nodes $v\in\mathcal{V}_A$, $q_A$ is the parameterized attention vector for node type $A$, $\beta_{P_i}$ can be interpreted as the relative importance of the metapath $P_i$ to nodes of type $A$, and $\mathbf{h}_v^{\mathcal{P}_A}$ represents the final node embedding of $v$, namely a weighted sum of all metapath-specific node vectors of $v$. By integrating multiple metapaths, MAGNN can learn the comprehensive semantics ingrained in the heterogeneous graph.

\subsection{\gnn\ Architecture}
\label{sec:preliminary:arch}

\begin{figure}[!htb]
\begin{center}
\includegraphics[width=0.75\columnwidth]{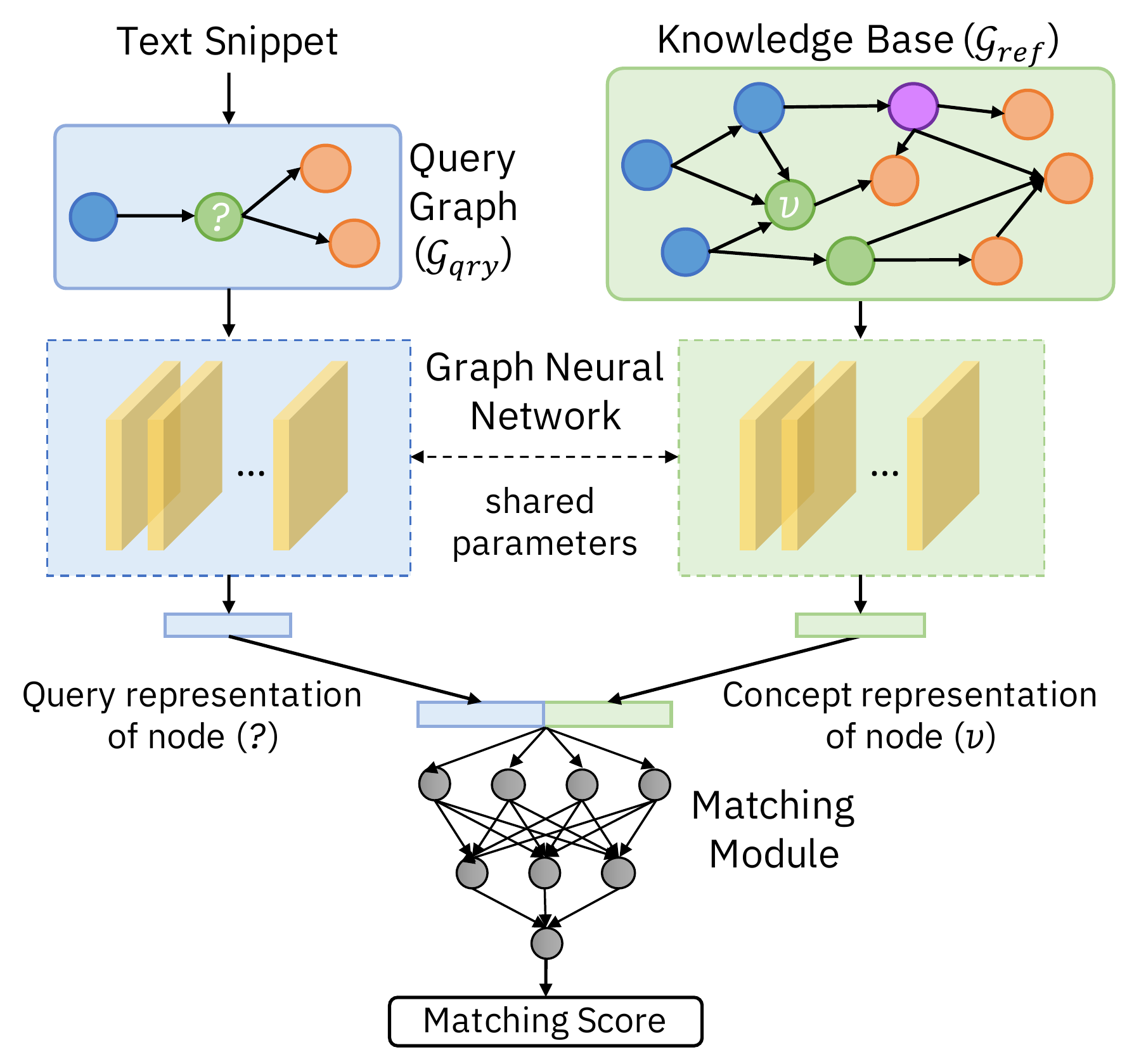}
\caption{\gnn\ architecture (best viewed in color).}
\label{fig:edgnn}
\end{center}
\end{figure}

We now present an overview of \gnn\ (depicted in Figure~\ref{fig:edgnn}) for medical entity disambiguation. The basic idea is to represent both a medical KB and a given text snippet as heterogeneous graphs $\mathcal{G}_{\it ref}$ and $\mathcal{G}_{\it qry}$, respectively. Following the property graph model~\cite{DBLP:conf/edbt/DasSPCB14}, we assume that nodes are associated with literal attributes in both $\mathcal{G}_{\it ref}$ and $\mathcal{G}_{\it qry}$, where nodes and edges have different types. In $\mathcal{G}_{\it ref}$, nodes correspond to medical entities and edges correspond to relationships between those entities. The entity mentions and extracted relations from the text snippets are represented as nodes and edges in $\mathcal{G}_{\it qry}$. Section~\ref{sec:opt:query} describes the optimized query graph modeling in further details.

Medical KBs are often curated and updated from text corpora in medical literature. The text snippets are collected from these text corpora as well. Hence, the neighboring structures of $\mathcal{G}_{\it ref}$ and $\mathcal{G}_{\it qry}$ are expected to be similar. Inspired by Siamese networks~\cite{DBLP:conf/icml/LiGDVK19}, \gnn\ uses two identical graph neural networks (one of GraphSAGE, R-GCN, or MAGNN) to generate the graph embeddings that encode all local structural information centered around the nodes in $\mathcal{G}_{\it qry}$ and $\mathcal{G}_{\it ref}$, respectively. These two GNNs share the same parameters (i.e., weight matrices) and consume a node list and an edge list from both $\mathcal{G}_{\it ref}$ and $\mathcal{G}_{\it qry}$, respectively. In a node list, each row contains a node id, its attribute features, and its type. In an edge list, each row has a source node id (head), a destination node id (tail), and the edge type. More details can be found in~\cite{wang2019dgl}.

{\bf Model Training.}
\gnn\ learns the representation of each node (node `$v$' in Figure~\ref{fig:edgnn}) in $\mathcal{G}_{\it ref}$ based on its k-hop or metapath-based neighbors and the representation of the query concept (node `$?$' in Figure~\ref{fig:edgnn}) in $\mathcal{G}_{\it qry}$. Such representation captures not only the node features, but also the topological structure of each node's neighborhood in $\mathcal{G}_{\it ref}$ or $\mathcal{G}_{\it qry}$. The matching module calculates their matching score, indicating the likelihood of two nodes matching each other. The matching module can be a multi-layer perceptron with one hidden layer, a log-bilinear model, or simply a dot product. We optimize the model weights by minimizing the following loss function through negative sampling:
\begin{equation}
\begin{aligned}
\mathcal{L} = & -\sum_{(u,v)\in \Omega} log (\sigma (\textbf{\textit{h}}^\top_u \textbf{\textit{h}}_v)) - \sum_{(u,v)\in \Omega^-} log(\sigma (\textbf{\textit{h}}^\top_u \textbf{\textit{h}}_{v'})),
\end{aligned}
\label{eq:loss}
\end{equation}
where $\sigma$($\cdot$) is the sigmoid function, $\Omega$ is the set of observed (positive) node pairs, and $\Omega^-$ is the set of negative node pairs sampled from all unobserved node pairs. In our entity disambiguation scenario, a positive node pair consists of one node representing an ambiguous entity in the text snippet and one node representing its corresponding matching node in the medical KB, respectively. By default, \gnn\ adopts uniform negative sampling by corrupting one node in the positive node pairs, due to its simplicity and efficiency. An optimized negative sampling strategy is introduced in Section~\ref{sec:opt:neg}. The above loss is the cross entropy of classifying the positive pair correctly.

\section{Optimizations in \gnn}
\label{sec:opt}

\subsection{Semantic Augmentation for Query Graph}
\label{sec:opt:query}

Our first optimization allows domain knowledge from the medical KB to be injected into the query graph $\mathcal{G}_{\it qry}$ through processing the text snippet to emphasize critical information for entity disambiguation. This processing step includes entity mention extraction and query graph construction.

\begin{figure}[!htb]
\begin{center}
\includegraphics[width=0.6\columnwidth]{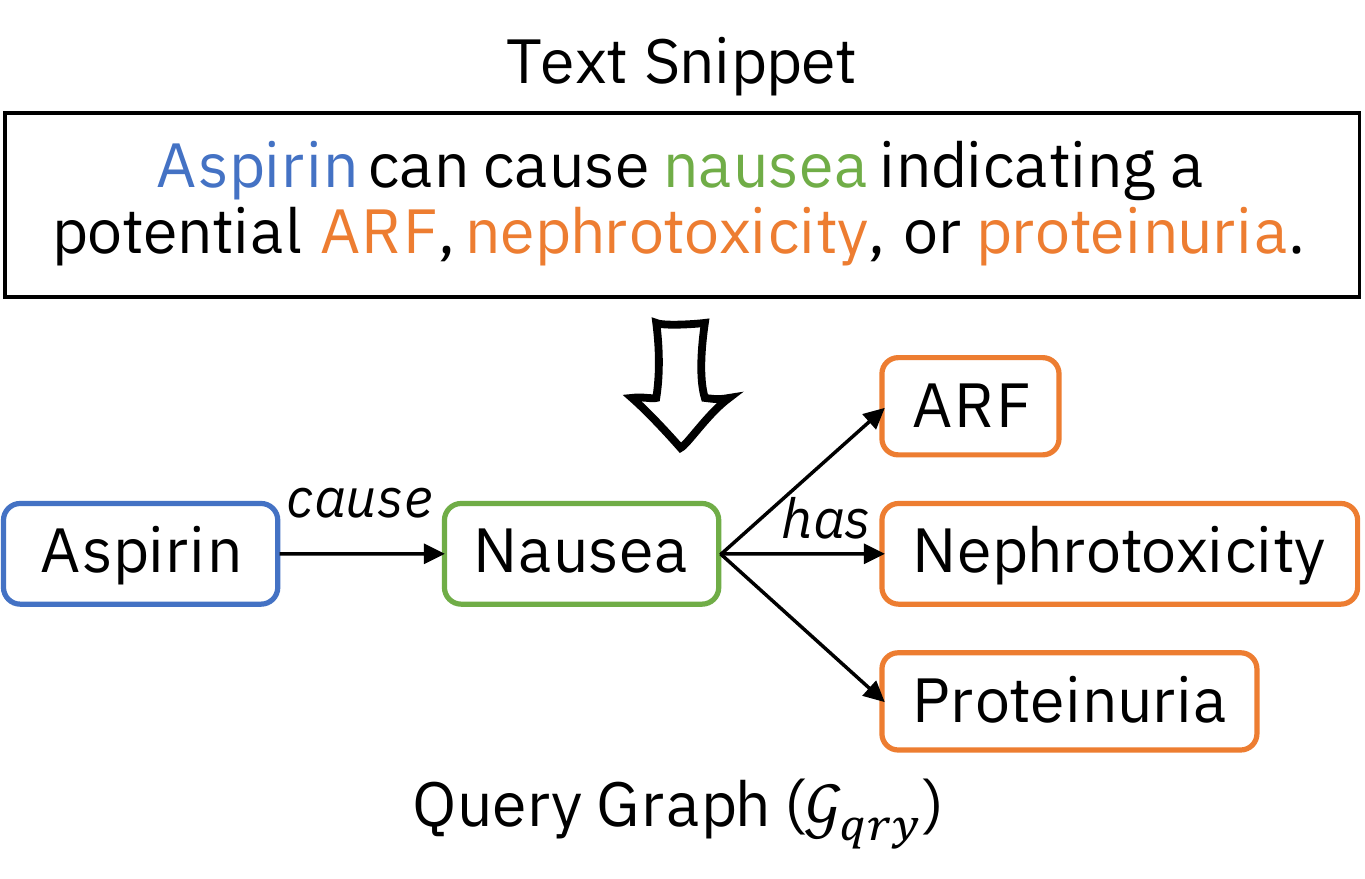} 
\caption{Text snippet to query graph (best viewed in color).}
\label{fig:queryGraph}
\end{center}
\end{figure}

\textbf{Augment Entity Mentions with Node Types from $\mathcal{G}_{\it ref}$.} 
To extract entity mentions from an input text snippet, i.e., named entity recognition (NER), many existing methods are available, including Stanford CoreNLP~\cite{snlp}, AllenNLP~\cite{Gardner2017AllenNLP}, and PyText~\cite{lample-etal-2016-neural}. In this work, we choose BioBERT~\cite{biobert}, a deep learning-based clinical NER model, fine-tuned on the medical KB. Consider the text snippet in Figure~\ref{fig:queryGraph}: \textit{``Aspirin can cause nausea indicating a potential ARF, nephrotoxicity, and proteinuria''}. In this sentence, we can identify the following terms as entity mentions of medical entities: ``\textit{Aspirin}'', \textit{``nausea''}, ``\textit{ARF}'', ``\textit{nephrotoxicity}'' and ``\textit{proteinuria}''.

Having entity mentions detected, we try to match them with the nodes in the medical knowledge base $\mathcal{G}_{\it ref}$. We exploit an inverted index of the entities in $\mathcal{G}_{\it ref}$ for the matching. Such inverted index includes not only the exact matches of these entities, but also synonyms, acronyms, and abbreviations of the entities in $\mathcal{G}_{\it ref}$. For the matched entity mentions, we further infer their entity types based on their corresponding entities in $\mathcal{G}_{\it ref}$. For example, we identify ``\textit{aspirin}'' as an instance of Drug, ``\textit{nausea}'' as an instance of AdverseEffect, and ``\textit{nephrotoxicity}'' as well as ``\textit{proteinuria}'' as instances of Finding in $\mathcal{G}_{\it ref}$. These identified entities can help us disambiguate the remaining entity mentions (e.g., ``\textit{ARF}''), for which a match is not found. Then, these identified entity mentions are used as the node set in the query graph $\mathcal{G}_{\it qry}$. It is possible that an entity mention has multiple matches in $\mathcal{G}_{\it ref}$. In this case, we associate all entity types of these matches to the entity mention.

\textbf{Augment Relationships in $\mathcal{G}_{\it qry}$.} 
One can create a query graph by connecting each node pair with an edge (self-loops are also added in this process)~\cite{cao2018neural,DBLP:conf/www/WuZMGSH20}. The resulting query graph can be considered as an undirected graph that describes the dependencies between entity mentions. However, such approach fails to utilize the domain knowledge from the medical knowledge base. Namely, the constructed query graph does not capture different relationships between a pair of entities, which provide critical contextual information to entity disambiguation.

To address this issue, we leverage the domain knowledge from $\mathcal{G}_{\it ref}$ to augment the query graph $\mathcal{G}_{\it qry}$. Specifically, we introduce an edge between a pair of nodes $u^q$ and $v^q$ (i.e., entity mentions) in $\mathcal{G}_{\it qry}$, if there exist two nodes $u^r$ and $v^r$ in $\mathcal{G}_{\it ref}$, such that $u^q$ matches $u^r$, $v^q$ matches $v^r$, and there exists an edge between $u^r$ and $v^r$ in $\mathcal{G}_{\it ref}$. The type of the newly added edge can be inferred from the corresponding edge in $\mathcal{G}_{\it ref}$ as well. To continue the above example, the nodes ``\textit{Aspirin}'' and ``\textit{nausea}'' are connected by an edge of type CAUSE in $\mathcal{G}_{\it ref}$ (shown in Figure~\ref{fig:hg}). Hence the newly added edge in $\mathcal{G}_{\it qry}$ is of type CAUSE as well. For those entity mentions (e.g., ``\textit{ARF}'') that do not have their matches in $\mathcal{G}_{\it ref}$, we rely on entity types obtained from NER to find the corresponding node type in $\mathcal{G}_{\it ref}$ and further identify the edges associated with the node type. Subsequently, we add an edge between the unknown entity and the existing entities if the corresponding node types are connected in $\mathcal{G}_{\it ref}$. This newly added edge in $\mathcal{G}_{\it qry}$ is also augmented with the corresponding edge type information from $\mathcal{G}_{\it ref}$. The overall query graph augmentation method is presented in Algorithm~\ref{algo:querygraph}. 

\begin{algorithm}[htb]
\begin{algorithmic}[1]
\Require A knowledge graph $\mathcal{G}_{\it ref}$, a text snippet $T$
\Ensure A query graph $\mathcal{G}_{\it qry} (\mathcal{V}_{\it qry}, \mathcal{E}_{\it qry})$
\State $\mathcal{G}_{\it qry}$ $\leftarrow$ $\emptyset$
\State \textit{EM} $\leftarrow$ NER($T$) $//$get all entity mentions
\State \textit{EM$_{\it match}$} $\leftarrow$ match($EM$, $\mathcal{G}_{\it ref}$) $//$get matching entity mentions
\State \textit{EM$_{\it unknown}$} $\leftarrow$ \textit{EM} $\setminus$ \textit{EM$_{\it match}$}
\State $\mathcal{V}_{\it qry}$.addNode(\textit{EM$_{\it match}$})
\ForAll {pair of nodes $u^q, v^q \in \mathcal{V}_{\it qry}$}
    \State $u^r\leftarrow$ \textit{EM$_{\it match}$}.getMatch($u^q$)
    \State $v^r\leftarrow$ \textit{EM$_{\it match}$}.getMatch($v^q$)
    \If {$e$ = ($u^r$,$v^r$) $\in \mathcal{E}_{\it ref}$}
        \State $\mathcal{E}_{\it qry}$.addEdge($u^q, v^q$, $e$.{\it type}) 
    \EndIf
\EndFor
\ForAll {$em \in$ \textit{EM}$_{\it unknown}$}
    \State \textit{et} $\leftarrow$ $em$.getEntityType()
    \State \textit{EdgeTypeSet} $\leftarrow$ $\mathcal{G}_{\it ref}$.getEdgeTypes(\textit{et})
    \State \textit{EntityTypeSet} $\leftarrow$ $\mathcal{G}_{\it ref}$.getEntity(\textit{EdgeTypeSet})
    \State $\mathcal{V}_{\it qry}$.addNode(\textit{em}) $//$add $em$ to $\mathcal{G}_{\it qry}$
    \State $u^q \leftarrow em$
    \ForAll {$v^q \in$ $\mathcal{V}_{\it qry}$ \textbf{and} $v^q \neq u^q$}
        \If {$v^q$.getEntityType() $\in$ \textit{EntityTypeSet}}
            \State \textit{edgeType} $\leftarrow$ \textit{EdgeTypeSet}.get($v^q$.getEntityType(), \textit{et})
            \State $\mathcal{E}_{\it qry}$.addEdge($u^q$, $v^q$, \textit{edgeType})
        \EndIf
    \EndFor    
\EndFor
\State \Return $\mathcal{G}_{\it qry}$
\end{algorithmic}
\caption{Query Graph Augmentation}
\label{algo:querygraph}
\end{algorithm}

\subsection{Semantic-Driven Negative Sampling}
\label{sec:opt:neg}

Negative sampling is used in our loss function (Equation~\ref{eq:loss}) as an approximation of the normalization factor of edge likelihood~\cite{DBLP:conf/nips/MikolovSCCD13}. Random negative sampling is commonly adopted in graph representation learning~\cite{DBLP:conf/nips/HamiltonYL17} due to its simplicity and efficiency. However, most of the negative samples are trivial cases from which the model does not gain much discriminative power~\cite{DBLP:conf/icde/ZhangYSC19}. Generative adversarial network (GAN), has been introduced in negative sampling~\cite{DBLP:conf/aaai/WangLP18} to avoid the problem of vanishing gradient and thus to obtain better performance. However, using GAN increases the number of training parameters and leads to instability and degeneracy~\cite{DBLP:conf/icde/ZhangYSC19}.

To solve the above issues, for every positive training example, we provide \textit{difficult} negative examples for our \gnn\ to learn. Intuitively, these negative examples are very close to the positive entity in the embedding space either due to their lexical or structural features. Hence, we generate them by utilizing two different sources of similarity evidence, which emphasize on both semantic and structural relatedness between positive and negative examples. 

{\bf Semantic Similarity.} Difficult negative examples should be semantically similar to the positive entity in $\mathcal{G}_{\it ref}$. For example, a positive node pair is (\textit{``MH''}, \textit{``Malignant hyperpyrexia''}), in which \textit{``MH''} is the ambiguous entity mention in $\mathcal{G}_{\it qry}$ and \textit{``Malignant hyperpyrexia''} is the labeled positive entity in $\mathcal{G}_{\it ref}$. Then, (\textit{``MH''}, \textit{``Malignant hyperthermia''}) can be considered as a difficult negative example since the semantic similarity between these two entities is very high. To find such negative examples, we reuse the initial node (i.e., entity) embeddings in $\mathcal{G}_{\it ref}$ and compute the cosine similarity between each positive example and other entities in $\mathcal{G}_{\it ref}$. Note that these initial node embeddings can be obtained using language models such as BERT~\cite{bert} on each node in both $\mathcal{G}_{\it qry}$ and $\mathcal{G}_{\it ref}$.

{\bf Structural Similarity.} Difficult negative examples should also share many common neighbors with the positive entity in $\mathcal{G}_{\it ref}$. Intuitively, two entities are similar if they are related to similar entities. Different graph similarity metrics are defined, ranging from graph edit distance (GED)~\cite{ged}, maximum common subgraph 
~\cite{mcs}, to graph kernels~\cite{graphkernel}. In this work, we choose the commonly used GED to compute the structural similarity of two entities in $\mathcal{G}_{\it ref}$. Only the local (i.e., 1-hop) neighbors of an entity are used in GED, which substantially reduces the computational cost. Our choice aligns well with the observation that 1-hop neighbors provide the most significant structural information in terms of a node representation. 

We integrate the above two measures into the scoring function: $sim = sim_{se} \cdot sim_{st}$, where $sim_{se}$ is the cosine similarity between two entity embeddings and $sim_{st}$ is the normalized GED according to~\cite{DBLP:conf/gbrpr/QureshiRC07}. The resulting similarity score is in the range of [0, 1]. Before training, negative examples are generated by ranking entities in $\mathcal{G}_{\it ref}$ according to their similarity scores with respect to the ambiguous entities in the labeled training set. The top-ranked examples are randomly sampled. As a result, the hard negative examples are more similar to the query than random negative examples, thus forcing the model to learn to disambiguate entities at a finer granularity. To reduce the computational cost, we only consider the immediate neighbors of an entity in the positive example as candidates for negative examples. These negative examples are guaranteed to be negative, since the KB is a complete graph (no missing nodes/edges) and only one entity matches the ambiguous mention. This is different from link prediction, where a missing positive link can be falsely selected as a negative example.

During training, we adopt a curriculum training scheme~\cite{pinsage} where \gnn\ will learn from easy negative examples first, but then gradually focus on difficult ones. Specifically, no difficult examples are used in the first epoch of training such that our \gnn\ can quickly find an area in the parameter space where the loss is relatively small. We then add difficult negative examples in subsequent epochs, focusing the model to learn how to disambiguate highly related entities from only slightly related ones.

\section{Experimental Evaluation}
\label{sec:exp}

\subsection{Datasets}
\label{sec:exp:dataset}

We use the following datasets from the medical domain as heterogeneous graphs to evaluate the performance of our method. Each dataset is used as a KB by itself. There is only one mention to be disambiguated in each text snippet, and the goal is to find its corresponding entity in the KB. Simple statistics of the KBs corresponding to these datasets are summarized in Table~\ref{tab:dataset}.

\begin{itemize}[leftmargin=*]
\item MDX is a medical KB\footnote{https://www.ibm.com/products/micromedex-with-watson} that contains information about drugs, adverse effects, indications, findings, etc. It is manually curated from medical literature by editorial staff, and the text snippets are extracted from the literature as well. The ground truth for MDX is provided by the editorial staff.

\item MIMIC-III~\cite{mimiciii} is a public data set containing 40,000 anonymized patient health-related records. It includes information such as demographics, laboratory test results, medications, and diagnoses.

\item Bio CDR~\cite{10.1093/database/baw068} consists of 1,500 PubMed abstracts annotated with mentions of chemicals, diseases, and relations between them.

\item NCBI~\cite{DOGAN20141} consists of 700 PubMed\footnote{https://pubmed.ncbi.nlm.nih.gov/} abstracts annotated with disease mentions and their corresponding concepts in MeSH\footnote{https://meshb.nlm.nih.gov/search}.

\item ShARe~\cite{pradhan-etal-2014-semeval} comprises 433 anonymized clinical notes (400 training and 133 test), obtained from the MIMIC II\footnote{https://archive.physionet.org/mimic2/} clinical dataset and annotated with disorder mentions.

\end{itemize}

In public datasets, ground truths are provided in the following form: \texttt{``Text'': ``A common human skin tumour is caused by activating mutations.'', ``Mentions'': [\{``mention'': ``skin tumor'',``start\_offset'':15, ``end\_offset'':26, ``category'':\\``Disease'', ``link\_id'':``C0037286''\}]}. In this case, \textit{skin tumor} is the ambiguous mention and its corresponding entity in the KB is \textit{neoplasm of the skin}, which is represented by the concept unique identifier \texttt{C0037286} in the medical ontologies (UMLS, MeSH, etc).

\begin{table}[ht]
\centering
\caption{Dataset statistics.}
\begin{tabular}{c|ccccc}
\newtoprule
\textbf{Dataset} & MDX & MIMIC-III & NCBI & ShARe & Bio CDR \\
\newmidrule
\# Nodes    & 35,028 & 22,642  & 753 & 1,719  & 1,082\\ 
\# Edges    & 74,621 & 284,542 & 1,845 & 12,731 & 2,857\\ 
\newbottomrule
\end{tabular}
\label{tab:dataset}
\end{table}

\begin{table*}[htbp]
\centering
{
\caption{Results of entity disambiguation on five datasets.} 
\begin{tabular}{c|ccc|ccc|ccc}
\newtoprule
{\bf Methods} &
\multicolumn{3}{c|}{DeepMatcher} &
\multicolumn{3}{c|}{NormCo} &
\multicolumn{3}{c}{NCEL}
\\
\newmidrule
{\bf Datasets} & 
{\bf Precision} & {\bf Recall} & {\bf F1} &
{\bf Precision} & {\bf Recall} & {\bf F1} &
{\bf Precision} & {\bf Recall} & {\bf F1} 
\\
\newmidrule
MDX         & 0.656	& 0.700	& 0.677	
            & 0.687 & 0.634	& 0.659	
            & 0.673	& 0.659	& 0.666 \\
MIMIC-III   & 0.708	& 0.567	& 0.630	
            & 0.747	& 0.692	& 0.718	
            & 0.716	& 0.624	& 0.667	\\
NCBI        & 0.783	& 0.815	& 0.799	
            & 0.863	& 0.818	& 0.840	
            & 0.816	& 0.793	& 0.804	\\
ShARe       & 0.694	& 0.639	& 0.665	
            & 0.726	& 0.623	& 0.671	
            & 0.753	& 0.631	& 0.687	\\
Bio CDR     & 0.837	& 0.816	& 0.826	
            & 0.866	& 0.805	& 0.834	
            & 0.857	& 0.829	& 0.843	\\
\newmidrule
\newmidrule
{\bf Methods} &
\multicolumn{3}{c|}{\gnn\ (GraphSAGE)} &
\multicolumn{3}{c|}{\gnn\ (R-GCN)} &
\multicolumn{3}{c}{\gnn\ (MAGNN)}
\\
\newmidrule
{\bf Datasets} & 
{\bf Precision} & {\bf Recall} & {\bf F1} &
{\bf Precision} & {\bf Recall} & {\bf F1} &
{\bf Precision} & {\bf Recall} & {\bf F1}
\\
\newmidrule
MDX         & 0.614 & 0.900	& 0.730	
            & 0.722	& 0.867	& 0.788	
            & \bf0.725	& \bf0.967 & \bf0.829 \\
MIMIC-III   & 0.786	& \bf0.733	& \bf0.759	
            & 0.810	& 0.567	& 0.667	
            & \bf0.826	& 0.633	& 0.717	\\
NCBI        & \bf0.924	& \bf0.856	& \bf0.889	
            & 0.912	& 0.823	& 0.865	
            & 0.915	& 0.861	& 0.887	\\
ShARe       & 0.794	& 0.829	& 0.811	
            & 0.806	& 0.833	& 0.819	
            & \bf0.825	& \bf0.879	& \bf0.851	\\
Bio CDR     & 0.853	& 0.845	& 0.849	
            & \bf0.896	& \bf0.867	& \bf0.881	
            & 0.864	& 0.853	& 0.858	\\
\newbottomrule
\end{tabular}
}
\label{tab:main}
\end{table*}

Each dataset is split into training (70\%), validation (15\%), and testing (15\%) sets unless otherwise stated. For NCBI, it is split into a training set of 500 abstracts, a validation and a test set of 100 abstracts each. For Bio CDR, it comes with a training set of 1000 and a test set of 500 abstracts. We further split its training set into a training and a validation set of 800 and 200 abstracts. For \gnn\ variants, we add the same number of negative node pairs described in Section~\ref{sec:opt:neg} to the validation and testing sets. These negative samples purposely cover different cases (e.g., abbreviation, synonym, acronym, and simplification).

\subsection{Systems}
\label{sec:exp:baseline}

We evaluate our approach \gnn\ using three different GNNs:  GraphSAGE~\cite{DBLP:conf/nips/HamiltonYL17}, R-GCN~\cite{DBLP:conf/esws/SchlichtkrullKB18}, and MAGNN~\cite{MAGNN}. We also compare \gnn\ with the state-of-the-art methods DeepMatcher~\cite{10.1145/3183713.3196926}, NormCo~\cite{Wright2019NormCoDD}, and NCEL~\cite{cao2018neural}, which are briefly described below.

\begin{itemize}[leftmargin=*]
    \item \gnn\ (GraphSAGE) employs GraphSAGE, designed for homogeneous graphs. It models the graph topology through neighbors aggregation on the node attributes.
    
    \item \gnn\ (R-GCN) leverages R-GCN, which handles different relationships between entities in a KB. It learns multiple convolution matrices corresponding to different edge types.
    
    \item \gnn\ (MAGNN) adopts MAGNN, which learns the representation of nodes based on their metapath-based neighbors with attention mechanisms at both node and semantic levels. 
    
    \item DeepMatcher is a supervised deep learning solution designed for entity resolution in a tabular setting. In our setting, an input to DeepMatcher is a tuple containing an ambiguous mention (e.g., \textit{skin tumor}) from a text snippet and an entity (e.g., \textit{neoplasm of the skin}) in the KB. We train and evaluate DeepMatcher with positive and negative tuples. Although the structural information from text snippets and KBs is not available to DeepMatcher, we choose it as an exemplar RNN method focusing on matching entities.

    \item NormCo uses a deep coherence model for disease entity normalization, which considers the semantics of an entity mention and the topical coherence of the mentions within a text snippet.
    
    \item NCEL creates a graph for candidates of mentions and then apply GCN to improve the disambiguation by directly aggregating information from linked nodes.
\end{itemize}

\textbf{Implementation Details.}
For the baseline systems (i.e., DeepMatcher, NormCo, and NCEL), we use the original hyper-parameter settings described in their papers, respectively. For DeepMatcher, we select its attention model since it has been shown effective on textual entity matching tasks in~\cite{10.1145/3183713.3196926}. For all \gnn\ variations, we employ the Adam~\cite{DBLP:journals/corr/KingmaB14} optimizer with the learning rate set to 0.001, the weight decay set to 0.001, and dropout rate to 0.5. We use the same splits of training, validation, and testing data sets for all models, and train the GNNs for 100 epochs and apply early stopping with a patience of 30. For \gnn\ using R-GCN and MAGNN, we set the dimension of the attention vector to 128. For \gnn\ using MAGNN, we set the number of attention heads to 2; we set the dimension of the attention vector in metapath aggregation to 128. For a fair comparison, we set the embedding dimension to 128 for all the above methods.

\begin{table*}[htbp]
\centering
{
\caption{Results of two optimization techniques on \gnn.} 
\begin{tabular}{c|c|ccc|ccc|ccc}
\newtoprule
\multirow{2}{*}{\bf Methods} & \multirow{2}{*}{\bf Datasets} &
\multicolumn{3}{c|}{Basic} &
\multicolumn{3}{c|}{Query graph augmentation} &
\multicolumn{3}{c}{Negative sampling}
\\
\cline{3-11}
 & & 
{\bf Precision} & {\bf Recall} & {\bf F1} & 
{\bf Precision} & {\bf Recall} & {\bf F1} & 
{\bf Precision} & {\bf Recall} & {\bf F1}
\\
\newmidrule
\multirow{2}{*}{\gnn\ (GraphSAGE)} & MIMIC-III
            & 0.747	& 0.702	& 0.724
            & 0.747	& 0.702	& 0.724
            & \bf0.786 & \bf0.733 & \bf0.759\\
& NCBI      & 0.869	& 0.821	& 0.844
            & 0.869	& 0.821	& 0.844
            & \bf0.924 & \bf0.856 & \bf0.889\\
\newmidrule
\gnn\ (R-GCN) & Bio CDR
            & 0.825	& 0.798	& 0.811
            & \bf0.863 & \bf0.826 & \bf0.844
            & 0.846	& 0.805	& 0.825\\
\newmidrule
\multirow{2}{*}{\gnn\ (MAGNN)} & MDX
            & 0.671	& 0.827	& 0.741
            & 0.694 & 0.863	& 0.769
            & \bf0.713 & \bf0.925 & \bf0.805\\
& ShARe     & 0.754	& 0.824	& 0.787
            & 0.796	& \bf0.868 & \bf0.830
            & \bf0.813 & 0.842 & 0.827\\
\newbottomrule
\end{tabular}
}
\label{tab:ablation}
\end{table*}

\subsection{Main Results}
\label{sec:exp:result}

We measure the performance of all methods using precision, recall, and F1, which are typical metrics for the evaluation of the entity disambiguation task~\cite{cao2018neural,Wright2019NormCoDD}. We report the average measurements of all methods on the test set for 100 repetitions. Table 3 reports the results of \gnn\ and other methods on all five datasets. The major findings are summarized as follows:

\begin{itemize}[leftmargin=*]
    \item Our \gnn\ variants consistently outperform other solutions in terms of precision, recall, and F1 on all datasets. The best performing \gnn\ variant offers an average improvement of 7.3\% in terms of F1 score, compared to the other best performing solutions. Among five datasets, we observe that all models perform better on NCBI and Bio CDR. The reason is that the graph complexity and semantic richness of NCBI and Bio CDR are simpler than the other datasets. The gain is much more significant on MDX (15.2\%) and ShARe (16.4\%) datasets. This fact manifests the expressive capability of our \gnn\ method to capture rich graph structures from both text snippets and KBs in medical entity disambiguation.
    
    \item Among all three \gnn\ variants, \gnn\ (MAGNN) achieves the highest average F1 score on all datasets, despite \gnn\ (GraphSAGE) and \gnn\ (R-GCN) achieve the best performance on MIMIC-III, NCBI, and Bio CDR datasets respectively. It is worth noting that \gnn\ (MAGNN) offers an average improvement of 2.1\% and 2.4\%, in terms of F1 score compared to \gnn\ (GraphSAGE) and \gnn\ (R-GCN), respectively. The results show that \gnn\ (MAGNN) captures both semantic and structural features by aggregating specific type of neighbors in the KBs, improving the performance of medical entity disambiguation. The other two \gnn\ variants deliver the best results on NCBI and Bio CDR datasets respectively as the complexities of these two datasets are less than the other ones. 
    
    \item Regarding the use of various graph features, DeepMatcher and NormCo only uses the text attributes of the compared entities, missing the opportunities to leverage more contextual information available in the graphs. NCEL incorporates GCN into its neural network to utilize only a subset of nodes next to the entity mentions but does not take edge types into consideration. \gnn\ (GraphSAGE) does not differentiate the contextual information aggregated via different edge types neither. This can be problematic when information gathered via certain edge types are not equally important. \gnn\ (R-GCN) tackles this issue by introducing an edge-aware aggregation function. \gnn\ (MAGNN) shows the expressive power provided by the metapath-based aggregation to explore the rich structural and semantic information in a KB, which eventually results in the best all-around performance.
\end{itemize}


\subsection{\gnn\ Model Studies}
\label{sec:exp:ablation}

{\bf Optimizations in \gnn.}
To make an ablation study on \gnn, we first evaluate the performance of our basic \gnn\ without two optimization techniques introduced in Section~\ref{sec:opt}, \gnn\ with semantic augmentation for query graph, and \gnn\ with semantic-driven negative sampling. For each dataset, we choose the best performing \gnn\ variant from Table 3. The major findings are summarized from Table 4. 

We observe that the semantic-driven negative sampling improves the basic \gnn\ (GraphSAGE) by 3.5\% and 4.5\% in terms of F1 score on MIMIC-III and NCBI, respectively. The query graph augmentation does not help at all in this case as GraphSAGE is not a relation-aware GNN. Similarly, \gnn\ (MAGNN) benefits more from the semantic-driven negative sampling strategy on MDX (+6.4\%). 
On the other hand, the query graph augmentation is more effective on BioCDR and ShARe datasets. Compared to the basic \gnn, the improvements are 3.3\% and 4.3\%, respectively. The reason is that the additional semantic information from the augmented query graph is more representative when the KB is simple.

These observations demonstrate that the query graph augmented with domain knowledge from the medical KB helps \gnn\ focus on the right structural information when making the matching decision. The semantic-driven negative sampling strategy, on the other hand, provides \gnn\ with harder examples, resulting in more discriminative power for entity disambiguation. Together, two optimization techniques improve the \gnn's disambiguation capability across a variety of medical datasets.

\begin{figure}[!htb]
\begin{center}
\subfigure[Visualization in MDX]{
\includegraphics[width=0.8\columnwidth]{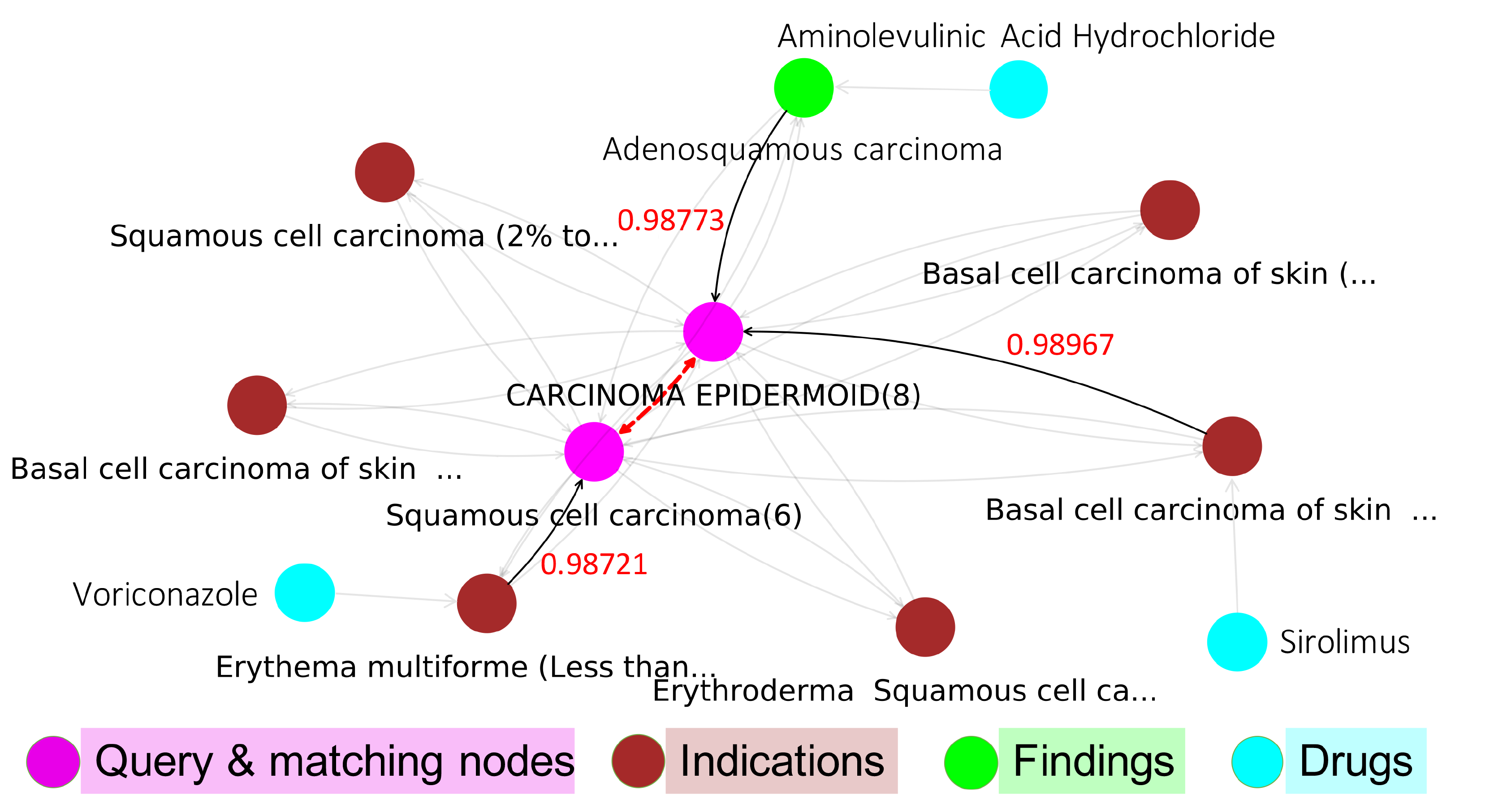}  
\label{fig:gnnexplainer}
}\\
\subfigure[Convergence]{
\includegraphics[width=0.8\columnwidth]{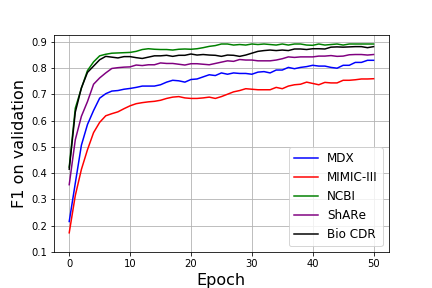}
\label{fig:convergence}
}
\caption{Model analysis (best viewed in color).}
\label{fig:model}
\end{center}
\end{figure}

Furthermore, we employ GNN-Explainer~\cite{DBLP:conf/nips/YingBYZL19} to visualize the important contributions of nodes and edges in KBs when finding the matching entity for the ambiguous mention. Due to the space constraint, we show one example using MDX dataset in Figure~\ref{fig:gnnexplainer}. GNN-Explainer highlights 3 most important (score range [0,1]) edges that contribute the most to matching ``squamous cell carcinoma'' with ``carcinoma epidermoid'' by \gnn. These edges carry critical information from different types of neighboring nodes, including ``adenosquamous carcinoma'' (Findings), ``basal cell carcinoma of skin'' (Indication), and ``erythema multiforme (less than 10\% epidermal detachment)'' (Indication). This indicates that our \gnn\ can learn and leverage the most semantically and structurally meaningful information among different types of entities and relations for entity disambiguation.

{\bf Convergence Analysis.} We analyze the convergence properties of \gnn, using the best performing ED-GNN variant from Table 3 for each dataset. The results, as shown in Figure~\ref{fig:convergence}, demonstrate that \gnn\ converges fast and achieves robust performance across all real-world datasets.

{\bf Number of Layers in \gnn.}
We also analyze the results of \gnn\ with 1 to 4 graph layers on all five datasets. Again, we choose the best performing ED-GNN variant from Table 3 for each dataset. In Table~\ref{tab:layer}, we observe that the optimal number of graph layers is 2 (for NCBI) or 3 (for MDX, MIMI-III, ShARe, and Bio CDR). When \gnn\ uses more than 3 layers, its performance declines. Although more layers allow \gnn\ to indirectly capture more distant neighborhood information by layer-to-layer propagation, such distant neighbors would introduce much noise and lead to more non-isomorphic neighborhood structures between the query graph and the KB.

\begin{table}[!htb]
\centering
\caption{Number of layers (F1).}
\begin{tabular}{c|ccccc}
\newtoprule
\textbf{\# layers} & MDX & MIMIC-III & NCBI & ShARe & Bio CDR \\
\newmidrule
1 & 0.691 & 0.641  & 0.815 & 0.731 & 0.785\\ 
2 & 0.751 & 0.704  & \textbf{0.891} & 0.825 & 0.843\\ 
3 & \textbf{0.829} & \textbf{0.759}  & 0.867 & \textbf{0.851} & \textbf{0.881}\\ 
4 & 0.743 & 0.727  & 0.831 & 0.806 & 0.829\\ 
\newbottomrule
\end{tabular}
\label{tab:layer}
\end{table}

\subsection{Error Analysis}
\label{sec:exp:case}

We also provide an error analysis on the entity mentions that are not disambiguated correctly by \gnn. Table~\ref{tab:error} breaks incorrect results in three categories below.

\begin{table}[!htb]
\small
\centering
\caption{Error analysis (\% of each test set).}
\begin{tabular}{c|>{\centering\arraybackslash}p{0.5cm}>{\centering\arraybackslash}p{1.2cm}>{\centering\arraybackslash}p{0.55cm}>{\centering\arraybackslash}p{0.6cm}>{\centering\arraybackslash}p{1.1cm}}
\newtoprule
\textbf{Error} & MDX & MIMIC-III & NCBI & ShARe & Bio CDR\\
\newmidrule
$\mathcal{G}_{\it qry}$ construction    & 9.5\% & 8.7\% & 1\% & 3.8\% & 2.2\%\\
Insufficient structure                  & 4.3\% & 9.8\% & 6\% & 3\% & 5.2\%\\
Highly similar nodes                    & 8\% & 4.8\% & 4\% & 3\% & 4.4\%\\
\newbottomrule
\end{tabular}
\label{tab:error}
\end{table}

\textbf{Query Graph Construction Error to $\mathcal{G}_{\it qry}$.} We observed that the semantic augmentation for query graph does not always lead to a correct query graph. The reasons are twofold. First, as described in Section~\ref{sec:opt:query}, an entity mention may be associated with multiple entity types. For example, \textit{``rash''} can be an instance of either Finding or AdverseEffect in MDX. Hence, the query graph may carry ambiguous semantic information that confuses \gnn. Second, multiple entity types can also lead to additional relationships in the query graph. These relationships could be irrelevant to the actual text snippet, causing \gnn\ to mismatch the ambiguous entity with incorrect entities in the KB.

\textbf{Insufficient Structural Information in $\mathcal{G}_{\it qry}$.} We observed that almost 50\% of the errors are due to a lack of graph structural information from text snippets. When a text snippet is short, the constructed query graph often contains few nodes and edges. For example, in a text snippet \textit{``Graft failure due to FSGS recurrence''} from MIMIC-III, \textit{``Graft failure''} is the only neighbor entity of \textit{``FSGS recurrence''}. In this case, \gnn\ does not have enough structural information to leverage, and has to primarily rely on the textual features of the ambiguous entity. Consequently, it fails to discover the corresponding entity in the KB's embedding space.

\textbf{Highly Similar Nodes in $\mathcal{G}_{\it ref}$.} At times, \gnn\ fails to identify the correct entity in the KB (e.g., MIMIC-III), even when the query graph is correctly constructed. In such cases, the entity corresponding to the ambiguous mention is often located in a highly dense area of the KB, where many semantically and structurally similar candidates exist. This essentially corresponds to the difficult negative examples described in Section~\ref{sec:opt:neg}. \gnn\ is not able to learn all possible negative examples through the semantic-driven negative sampling.

\section{Related Work}
\label{sec:related}

\textbf{Graph Neural Networks.} Graph representation learning has been shown to be extremely effective, achieving promising results in various domains over graph-structured data~\cite{DBLP:journals/corr/LiTBZ15,DBLP:conf/iclr/KipfW17,DBLP:conf/nips/HamiltonYL17,DBLP:conf/iclr/VelickovicCCRLB18,DBLP:journals/corr/abs-1804-00823}. GCN~\cite{DBLP:conf/iclr/KipfW17} is a graph convolutional network via a localized first-order approximation of spectral graph convolutions. The seminal GNN framework, GraphSAGE~\cite{DBLP:conf/nips/HamiltonYL17}, learns node embeddings through aggregating from a node's local neighborhood using inductive learning. Graph attention networks (GAT)~\cite{DBLP:conf/iclr/VelickovicCCRLB18} are introduced to learn the importance between nodes and their neighbors, and fuse the neighbors to perform node classification. 

Heterogeneous graph embedding has also received much research attention recently~\cite{DBLP:conf/esws/SchlichtkrullKB18,HAN,GATNE,MAGNN}, as many KBs also fall under the general umbrella of heterogeneous graphs. For example, R-GCN~\cite{DBLP:conf/esws/SchlichtkrullKB18} distinguishes different neighbors with relation-specific weight matrices. Heterogeneous graph attention network (HAN)~\cite{HAN} leverages a graph attention network architecture to aggregate information from the neighbors and then to combine various metapaths through the attention mechanism. Inspired by HAN, HetGNN~\cite{HetGAN} encodes the content of each node into a vector and then adopts a node type-aware aggregation function to collect information from the neighbors. HetGNN also uses attention over the node types of the neighborhood node to get the final embedding. MAGNN~\cite{MAGNN} captures all neighbor nodes and the metapath context using both intra-metapath aggregation and inter-metapath aggregation. Thus, the generated node embeddings preserve the comprehensive semantics in the heterogeneous graphs.

\textbf{Entity Disambiguation.} 
For many years, entity disambiguation (also referred to as entity linking) has been an active field of research~\cite{DBLP:journals/tkde/ShenWH15}. A related task, entity matching, has also been studied extensively in the context of structured data~\cite{DBLP:journals/dke/KopckeR10,DBLP:books/daglib/0030287}. Recently, \cite{10.1145/3183713.3196926,DBLP:conf/sigmod/GovindKCMNLSMBZ19} investigated various DL-based methods for entity matching, and concluded that although DL-based techniques do not offer significant advantages for structured data, they outperform current solutions~\cite{DBLP:conf/sigmod/GovindKCMNLSMBZ19} considerably for textual entity matching.
DoSeR~\cite{DBLP:conf/esws/ZwicklbauerSG16} relies on an RDF KB embedding~\cite{DBLP:journals/semweb/RistoskiRNLP19} for KB entities using known entity links to model the context in which those entities are mentioned in the text, which can subsequently be used to predict further mentions of such entities based on the mention’s context. NCEL~\cite{cao2018neural} applies graph convolutional network to integrate both local contextual features and global coherence information for entity linking. However, it only considers the immediate neighbors of an entity mention and does not take edge types into consideration. COM-AID~\cite{Dai:2018:FCL:3183713.3196907} introduces a composite attentional encode-decode neural network in healthcare. It encodes a concept into a vector and decodes the vector into a text snippet with the help of textual and structural contexts. NormCo~\cite{Wright2019NormCoDD} is designed for disease normalization. It models entity mentions using a semantic model, which consists of an entity phrase model using word embeddings and a coherence model of other disease mentions using an RNN. The final model combines both sub-models trained jointly. Unlike existing works in the field, we introduce a simple architecture that leverages state-of-the-art GNNs to encode the latent graph structure of the KB and the input text snippets for medical entity disambiguation.


\section{Conclusion}
\label{sec:conclusion}

In this paper, we study the entity disambiguation problem which plays an important role in medical knowledge graph curation and maintenance processes. We present \gnn, a medical entity disambiguation system, based on GNNs. \gnn\ uses a simple architecture to leverage state-of-the-art GNNs, and is further optimized by augmenting the query graph with domain knowledge from the medical KB as well as an effective negative sampling scheme to improve the disambiguation capability. The experimental results on multiple real-world medical KBs demonstrate that \gnn\ is effective and outperforms the state-of-the-art solutions.

\balance
\bibliographystyle{abbrv}
\bibliography{embedding}  

\end{document}